# Achieving Operational Scalability Using Razee Continuous Deployment Model and Kubernetes Operators

Srini Bhagavan
IBM Data and AI
Leawood, Kansas, USA
srinib@us.ibm.com

Saravanan Balasubramanian
IBM Data and AI
Littleton, Massachusetts, USA
saravanan.b@ibm.com

Prasad Reddy Annem
IBM Data and AI
Leawood, Kansas, USA
preddy@us.ibm.com

Thuan Ngo
IBM Data and AI
San Jose, California, USA
ngot@us.ibm.com

Arun Soundararaj
IBM Data and AI
San Jose, California, USA
arun.soundararaj@ibm.com

## ABSTRACT

It is safe to say the cloud era has arrived and for here to stay. The *COVID-19* pandemic has all but cemented that choice for us. Recent advancements in the cloud computing domain have resulted in huge strides toward simplifying the procurement of hardware and software for diverse needs. As-a-Service delivery models are making consumers rethink their Information Architecture and application deployment models. By moving their enterprise workloads to managed cloud offerings (private, public, hybrid), customers are delegating mundane tasks and labor-intensive maintenance activities related to network connectivity, procurement of cloud resource, application deployment, software patches, and upgrades, etc., This often translates to benefits such as high availability and reduced cost. The popularity of container and micro-services-based deployment has made Kubernetes [1] the defacto standard to deliver applications. However, even with Kubernetes orchestration, cloud service providers frequently have operational scalability issues due to lack of Continuous Integration and Continuous Deployment (CICD) [20] automation and increased demand for human operators when managing a large number of software deployments across multiple data centers/availability zones. Kubernetes solves this in a novel way by creating and managing custom applications using Operators [10]. Agile methodology advocates incremental CICD which are adopted by cloud providers. However, ironically, it is this same continuous delivery feature of application updates, Kubernetes cluster upgrades, etc., that is also a bane to cloud providers. In this paper, we will demonstrate the use of IBM open-source project Razee as a scalable continuous deployment framework to deploy open-source RStudio [22] and Nginx Operators [19]. We will discuss how IBM Watson SaaS application Operator [12], Blockchain applications [7], and Kubernetes resources updates, etc., can be deployed similarly and the use of Operators to perform application life cycle management. We assert that using Razee [5] in conjunction with Operators on Kubernetes simplifies application life cycle management and increases scalability.

## 1 Introduction

Gartner estimates that *by 2022 worldwide public cloud services market is estimated to grow to ~$354.6 billion ~60% of organizations will use an external service provider's cloud-managed service offering [6]*. Cloud computing is the ability to procure services and assets on demand by the users without requiring them to perform active management. Managed services preclude the user from worrying about daily mundane tasks. Furthermore, the task of infrastructure, IT specialists to maintain and update software, troubleshoot down systems, etc., are the tasks of the managed service cloud provider. This allows the customer to focus and optimize their business operations rather than staff up and operate their IT infrastructure. Enterprise customers have to meet regulatory requirements, *FedRAMP [24], SOC2 [25], HIPPA [26],* and other compliance specifications to market their offerings. Over recent years, cloud providers are certifying their cloud stack to be compliant, which has attracted traditional on-premises customers to the cloud. Additionally, Service Level Agreements (SLA) are published and maintained by these providers for each of the services they offer. The benefits of operating their business in the cloud far outweigh the notion of giving up control. Cloud computing is here to stay and will continue to evolve.

An increased velocity of cloud adoption presents new challenges to cloud providers. The adoption of container technology, microservices architecture, and Kubernetes as the platform has helped to solve some of the scalability challenges. In a microservice architecture, an application is decomposed into many smaller components, and each component has its own responsibilities. Kubernetes is a container orchestration platform that helps to orchestrate microservices. Cloud providers have entrenched the use of Kubernetes (or similar container orchestration technologies) as their cloud application deployment strategy. Cloud SaaS offerings are being re-architected to become containerized to be able to deploy on Kubernetes platforms.

By using Kubernetes and other tools as part of the cloud ecosystem, there is a challenge presented related to the deployment of services into the Kubernetes cluster. DevOps infrastructure automation needs to be enhanced to handle continuous deployment of Kubernetes resources into the cluster, monitoring, and maintenance of the application stack that requires minimal human intervention. For example, if we use Jenkins [27] as a deployment tool in the Multi-Cluster Management (MCM) [16] environment, we must have a complex infrastructure built like shown in Figure 1. To deploy a Jenkin's service to multiple clusters, we first deploy a new Jenkins instance in the cloud provider environment. Jenkins uses the MCM controller to deploy services to multiple clusters. Finally, based on policies defined, MCM will target clusters to deploy application services. As we see we need to define these MCM controllers and policies and constantly update these policies and maintain this complex infrastructure just to have the services deployed on to multiple clusters. We also need to maintain a separate architecture flow defined to distinguish between the cluster service updates and application service deployments. Human interventions in form of the Site Reliability Engineers (SRE) to diagnose problems for down systems. Furthermore, with Jenkins, a separate infrastructure needs to be configured and maintained to handle monitoring operations.

The architecture proposed in this paper employs the use of Razee, which is an open-source continuous deployment tool. Razee, along with the notion of Kubernetes Operators solves the problems faced by Jenkins elegantly. Kubernetes Operators act as reliable SREs. They are designed to handle upgrades seamlessly, react to failures automatically and independently without many human interventions. The other problem Razee solves is that it has unique ways to monitor the Kubernetes resources deployed in the target clusters. All the application clusters would report their Kubernetes resource metadata information and are accessed via *RazeedashUI*, which provides the ability to configure alerts.

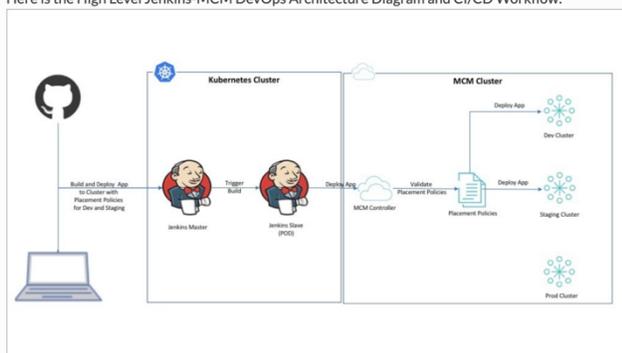

Figure 1: High-level architecture of Jenkins CICD workflow for Multi-Cloud Management. [16]

In the following sections, we will elaborate on the setup of Razee and its components onto the target clusters. We will deploy a sample Nginx service operator into the target clusters. In addition, we will discuss updating Operators as part of continuous deployment and understand how Razee automatically updates Kubernetes Operators, application instances, and deploys them onto multiple target clusters that are registered with Razee.

## 2 Background

In this section, we will have a brief discussion on the technologies used in this article.

### 2.1 Kubernetes

Kubernetes is an open-source platform for orchestrating containerized workloads and services. A Kubernetes cluster typically consists of one or more hosts working together to achieve certain goals which are typically not possible on standalone hosts due to lack of computing power or vertical salability limitations. Kubernetes provides a wide range of benefits such as automation of application deployments, scaling, and portability which makes Kubernetes a very attractive option to host workloads. The workloads and services are hosted on the cluster as one or more smallest deployable units of computing called pods. Each pod contains one or more working containers. Kubernetes clusters are logically divided into two planes, namely the control plane and the data plane. While the control plane is responsible for managing the cluster and the workload hosted on the cluster, the data plane is where the actual workload containers run. In the monolithic architecture, all components of the application are deployed together as one processing unit. This architecture makes it difficult to manage, scale, and upgrade the application. In contrast, the micro-service architecture lets us decompose the application into smaller components and allows us to deploy and update irrespective of the other components of the application. However, it raises the requirement to orchestrate these micro-services together. This is where Kubernetes steps in to orchestrate and deliver multiple micro-services into a single logical application.

### 2.2 Continuous Deployment

Continuous deployment [20] (CD) is a software development practice where software functionalities are deployed frequently through automated deployment to production systems. In this approach, software changes are typically incremental updates. It reduces risks made to the production system by delivery of incremental changes and not compromise the entire infrastructure with massive updates in a single upgrade. CD is a critical agile requirement for SaaS offerings. As the cloud providers operate on a vast number of application deployments across multiple clusters and data centers, they must employ a CD framework that operates at scale.

### 2.3 Razee an open-source CD tool

Razee is an open-source project developed by IBM designed to be used as a multi-cluster continuous delivery tool for Kubernetes. Razee employs the Kubernetes Operator pattern and uses a pull-based deployment model to apply Kubernetes artifacts on the target cluster. Razee supports grouping of clusters using labels and thus allows the Operator to roll out feature updates only to the selective cluster groups. Razee architecture consists of four main components namely *Razeedash, Razeedeploy, watchkeeper, and Razeedeployables* as shown in Figure 2.

### 2.4 Red Hat OpenShift Container Platform

Red Hat OpenShift [4] is an open-source hybrid cloud Kubernetes container application platform for enterprise application development and deployment. Red Hat OpenShift provides other platform services on top of Kubernetes such as in-built container registry, monitoring using Prometheus, and CICD pipeline using Jenkins. In other words, Kubernetes can be thought of as a kernel, and OpenShift is one of the distributions.

### 2.5 IBM Red Hat OpenShift Kubernetes Service

Red Hat OpenShift Kubernetes Service [28] is the managed Red Hat OpenShift on IBM Cloud. IBM manages the OpenShift Container Platform (OCP) for the customers. As the customer doesn't need to worry about managing the OCP cluster and the underlying infrastructure, the customer can focus on other core tasks, such as deploying and delivering the services and application workloads on the ROKS cluster.

### 2.6 Kubernetes Operators

Kubernetes Operators are akin to a coded version of Site Reliability Engineers (SRE) which is aware of exactly the operations that are part of a specific application. These Operators are coded with deep knowledge of specific instructions related to the operation of the application as part of deploy, update, scale automatically, maintain uptime availability, take data backups, etc., Operators enable users to create and manage custom applications by extending the Kubernetes API. Custom applications are defined as new resources using Custom Resource Definition by extending Kubernetes API rather than adding a new API server. Controllers of CRDs are custom applications that are responsible for running the reconciliation loops (current-to-desired application state) and manage custom applications. For example, when a pod goes down, the controller will not only spawn a new pod but also take care of any prerequisite steps that need to be run like electing and selecting a master among the running pods, which may not be possible with Kubernetes provided vanilla reconciliation.

## 3 Experience with Kubernetes Operator and Razee as Continuous deployment tool

In this section, we attempt to set up a continuous deployment model for the Kubernetes Nginx Operator using Razee, and later deploy and manage the lifecycle of an Nginx instance in the cluster using the Nginx Operator. At the end of the section, we will be sharing our findings and learning as Pros and Cons of Razee being used as a Continuous Deployment tool.

To setup Razee, we need the following components deployed:

a. **Razee Deploy components** - Razee Deploy components [5] are deployed in the target customers clusters where the SaaS services like Watson Studio, Blockchain, Nginx services need to be deployed as part of the continuous delivery process. Deploy components consist of Custom Resource Definitions (CRDs) and Controllers to manage the services deployments.
b. **Watch Keeper** - Watch keeper runs on the target customers clusters and is responsible for sending the state of the Kubernetes resources which are labelled to be monitored to Razeedash. We can add the watch keeper labels like `razee/watch-resource: "lite"` into the Kubernetes resources. Here 'lite' means collecting basic metadata and status information to show up on the Razee dashboard. The other options could be "detail", "debug", the latter two options will send detailed information on the Kubernetes resources to RazeeDash.
c. **Razee Dash components** - Razee dash components consist of Razeedash API server, MongoDB, and a UI component. The API server and MongoDB are responsible for maintaining the list of all the clusters that are registered with Razeedash and the monitoring data sent from those registered customers clusters via the watchkeepers.
d. **RazeeDeployables** - With RazeeDeployables, we can control and automate the rollout of your Kubernetes resources across multiple remote clusters. To achieve that, we need to add the target clusters to the Razeedash inventory and subscribe to the subscription channels hosted on the Razeedash. To subscribe, the target cluster needs to be configured with tags that are used in the Subscription channels in Razeedash. Then the target cluster will pull the Kubernetes artifacts from the subscription channel and apply those changes in the cluster.

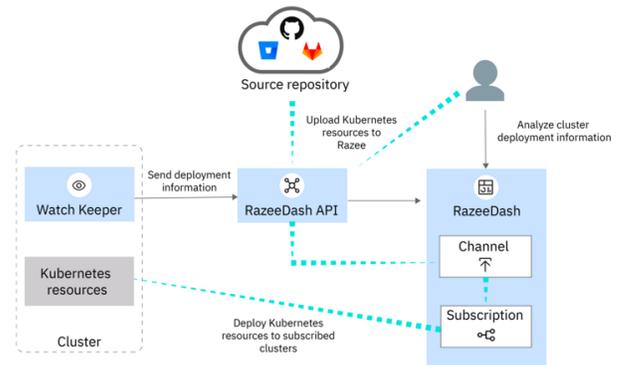

Figure 2: Razee Continuous Deployment framework 2020. [5]

### 3.1 Setting up Razee

We follow the documentation on Razee [5] GitHub page to set up the environment.

![Figure 3 image]

Figure 3: Custom Resource Definitions created in the target cluster

![Figure 4 image]

Figure 4: Razee Kubernetes resources deployed in the target cluster - Part A

Here are the design choices we made while setting up the working environment for Razee:

- To store the monitoring data sent by the watchkeepers of various clusters and to store the Kubernetes artifacts uploaded by us, we experimented using the default MongoDB that Razee deploys and as well as a cloud object storage instead of MongoDB.
- We then integrate with GitHub OAuth for authentication and authorization to access the Razeedash UI.

Figure 5: Razee Kubernetes resources deployed in the target cluster - Part B

- After the server cluster is set up with Razeedash, we can see the resources pertaining to Razee components would be deployed into the cluster as shown in Figures 3, 4,5. Figure 3, 4 shows all the Razee deployments resources and pods pertaining to it. Figure 5 shows the configmaps and secrets that got created. we can install Watch Keeper in every customer cluster that we want to monitor. The cluster where we install Watch Keeper can be a different cluster than the one where we installed RazeeDash. Typically, we will deploy the watch keeper components into the target clusters where we want to deploy the applications or Services (*Watson Studio, Nginx, Blockchain, etc*.,)
- With Watch Keeper set up in our customer cluster, we can retrieve deployment information and status from the target cluster and access, monitor, and analyze this data on Razeedash that we set up in the server cluster.
- The monitoring is achieved by adding the razee/watch-resource label as shown in Figure 6, to the labels section of all Kubernetes resources.

Figure 6: Razee labelled Kubernetes resource

- After we had added the label to our resources, Watch Keeper automatically scans our resource and sends data to the Razeedash API server that is displayed in RazeeDash UI as shown in Figure 7. In addition, Watch Keeper adds a Kubernetes event watcher to your resource so that Watch Keeper is notified by Kubernetes when the configuration of your resource changes.

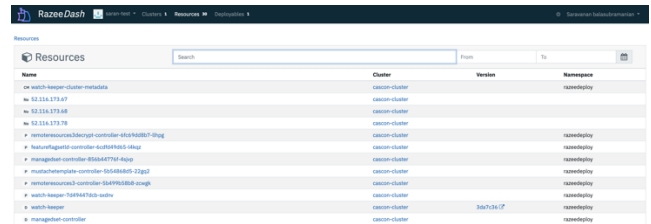

Figure 7: Razeedash UI Showing Kubernetes resource on a registered cluster

### 3.2 Deploy Nginx Operator using Razee

Now that we have Razeedash deployed on the server cluster and Razeedeploy and deployable components deployed on all the target clusters, we attempt to deploy Nginx Operator to the target cluster using the continuous deployment model and use that Operator to deploy instances of Nginx.

#### 3.2.1 Nginx Operator [14] deployment flow

1. The application administrator uses the Razeedash UI to create a channel for the Nginx Operator. After the channel is created, the admin can upload the initial version (1.0) of the Kubernetes YAML file with a version number that helps deploying the Nginx Operator.
2. Steps to create an Nginx Operator channel are as follows:

   - Figure 8 depicts the overall architecture of Nginx, RStudio, Blockchain Operators.
   - We need to create a new channel in the UI as shown in Figure 9 by clicking add button. Provide a name for the new channel. We used this same name '*nginx-operator*', in the curl command below:

     *curl --request POST --url "https://app.razee.io/api/v1/channels/nginx-operator/version"* --header "content-type: text/yaml" --header "razee-org-key: <ORG_API_KEY>" --header "resource-name: 1.0" --header "x-api-key: <RAZEE_API_KEY>" --header "x-user-id: <USER_ID>" --data-binary @all-in-one.yaml

     {"status":"success","version": {"uuid":"<uuid>","name":"1.0","location":"mongo"}}

   - The Operator channel looks like below, after successful channel creation we can see the version column is updated with the latest version.

Achieving Operational Scalability using Kubernetes Operators and
Razee Continuous Deployment Model

3. The application administrator uses the Razeedash UI to create a subscription for the Nginx Operator channel with tag (demo) to serves the initial version uploaded (1.0)

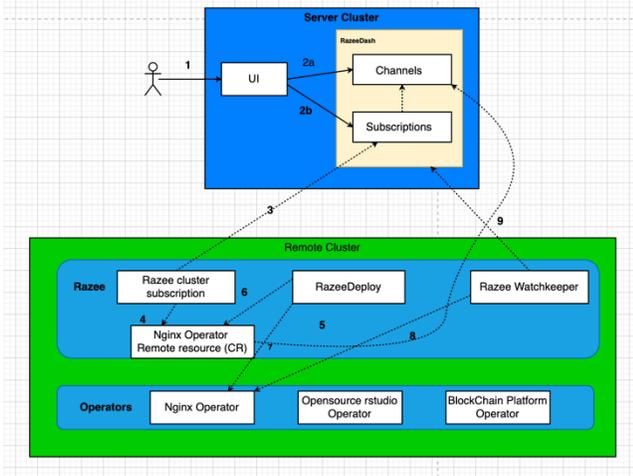

Figure 8: Nginx, RStudio, Blockchain Operators deployment model

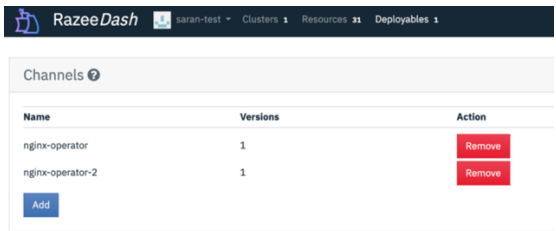

Figure 9: Razeedash showing the Channels and number of deployable versions of Kubernetes artifacts in the channel

4. Now, the target clusters which want to subscribe to the Nginx Operator 1.0 will subscribe by configuring the Clustersubscription configmap in the Razeedeploy namespace by adding the tags created in step2 as shown in Figure 10. Clustersubscription pod that is running on the target cluster is notified by the RazeeDash that there are subscriptions available for the configured tags and the repository URL of the artifacts in the subscription channel where it can be downloaded.

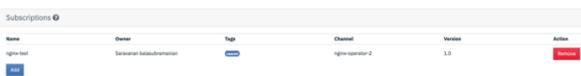

Figure 10: Razeedash showing the Subscription channel nginx-test with nginx-operator-2 channel with tag 'demo'

Figure 11: Command line showing the Nginx Operator resources created in the target cluster which has subscribed to subscription channel with label 'demo'

5. After artifacts are pulled, the Clustersubscription pod will create a custom resource (Razee Remote Resource) with information of the artifact repository URL.
6. The Custom resource (Remote resource) pointing to the location where the artifacts are stored.
7. Razeedeploy component (Remote Resource Controller) is notified of the creation of CR in step4.
8. Remote Resource Controller pulls the Kubernetes artifacts and applies the artifact in the cluster to deploy the Operator. When the Operator is deployed, all the necessary Kubernetes resources as shown in Figure 11 are created along with it such as Application CRD, Service account, Role, Role binding, service, and Operator pods.
9. The Operator deployment is labelled with razee labels and thus Razee watch keeper keeps track of the deployment and the status.
10. On a regular interval, the monitoring data of the labelled Kubernetes resources will be sent to the RazeeDash API Server and cached in the MongoDB, which can be viewed from the Razeedash UI.

**3.2.2 Nginx Instance deployment flow**

1. Figure 12 shows the overall Nginx Operator instance deployment architecture flow.
2. The application administrator directly interacts with the Kubernetes API server to create a custom resource of type Nginx.
3. Nginx Custom resource is created.
4. Nginx Operator listens to events of custom resource of type Nginx as shown in Figure 13.
5. Nginx Operator creates all the Kubernetes resources that are required for the deployment such as service, deployment, pod, and routes.
6. After the deployment is successful, the Nginx route when pasted on the browser will display the UI as shown in Figure 14, which ensures the proper functioning of the Nginx instance.
7. It also the responsibility of the Operator to run the reconciliation loop and keep the Nginx deployment in the desired state and thus maintaining the life cycle of the Nginx custom application.
8. When the created resources are labelled, the watch keeper will collect the data of the labelled Kubernetes and send it back to the RazeeDash API server, which can be viewed/monitored from RazeeDash UI.

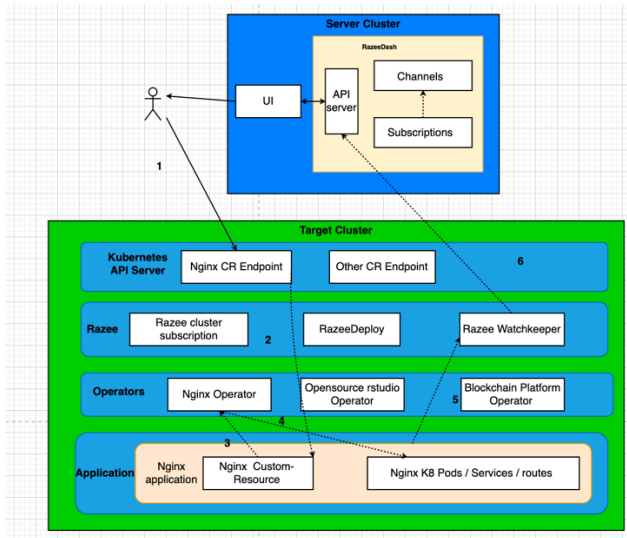

Figure 12: Nginx Instance deployment flow

**Curl command used to create the custom resource:**

*curl -k -X POST -d @nginx-cr.json -H "Authorization: <BEARER_TOKEN_OF_THE_TARGET_CLUSTER>" -H 'Accept: application/json' -H 'Content-Type: application/json' <CLUSTER_API_URL>/apis/example.com/v1alpha1/namespaces/<namespace>/Nginxes*

9. When the deployment is no longer required, deleting the Nginx custom resource will invoke the Operator to clean up the application and its attached resources.

**Curl command used to delete the custom resource:**

*curl -X DELETE -H "Authorization: <BEARER_TOKEN>" "<CLUSTER_URL>/apis/example.com/v1alpha1/namespaces/<namespace>/Nginxes/example-nginx "*

```
{
    "apiVersion": "example.com/v1alpha1",
    "kind": "Nginx",
    "metadata": {
        "name": "example-nginx"
    },
    "spec": {
        "replicaCount": 1,
        "ingress": {
            "enabled": true
        }
    }
}
```

Figure 13: Custom resource artifact of resource type Nginx

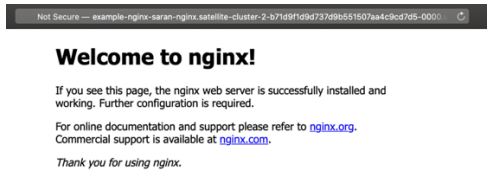

Figure 14: The webpage hosted by Nginx instance deployed using razee

### 3.2.3 Application update rollout flow

As discussed earlier in the paper, adopting agile methodology means evolutionary improvements and continuous delivery of those improvements to the consumers. For the application administrators, razee makes the rolling out updates as easy as flipping a switch. When the code changes or features needed to be released, the Continuous Integration (CI) pipeline builds a new docker application image and subsequently builds a new Operator image to use a new application image for deployments and updates. By creating a new Operator image, we can trigger a CI pipeline to create a new version of the YAML file incorporating the new changes and upload it to the razee channel. Now, we have Nginx Operator 2.0 deployment YAML in the razee channel as shown in Figure 15.

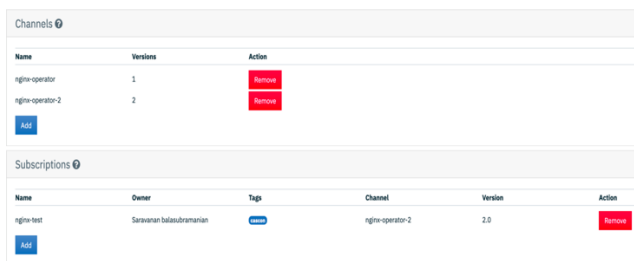

Figure 15: RazeeDash UI showing the Nginx Operator version changed from 1.0 to 2.0

To roll out application updates to a new version, the Operator subscription version needs to be changed from 1.0 to 2.0. On switching the version, the Clustersubscription pod running in the target clusters listens to the configured tag. On pulling the YAML, the Remote resource controller will run the reconciliation loop and will update the Operator from version 1.0 to 2.0. Subsequently, the Operator runs the reconciliation loop and updates all the applications running in that cluster automatically without any human interaction. By employing the pull method, the model scales very well compared to the conventional push model when dealing with many clusters. As the model does not differentiate where Kubernetes clusters are deployed, the model can be extended to any clusters deployed by any cloud provides or hybrid clouds. Figure 16 shows the welcome message loaded from version 2.0 of the Nginx Operator.

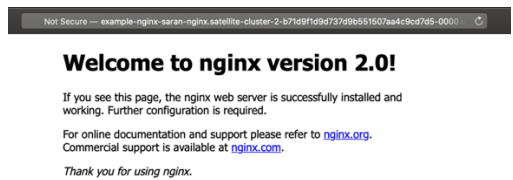

Figure 16: The webpage hosted by Nginx instance with updated UI.

### 3.2.4 Extending the model to other applications

We've demonstrated how Razee and Kubernetes can be used to deploy Nginx. A similar deployment topology can be used to deploy other more complex services such as open-source RStudio, Blockchain platform, and other applications with a Kubernetes operator.

Here, we used the OpenShift SDK to create operators for Open-source RStudio, Notebook, and Nginx. The SDK creates the folder structure as shown in Figure 17. The Operators are helm-based, and the helm chart of the application goes directly into the root of the folder created by the SDK. We made changes to the CRD in the file example.com_rstudios_crd.yaml as needed as shown in Figure 18 to extend the Kubernetes API to accommodate our RStudio custom application. After making the changes to the Helm chart to deploy the RStudio application, we used the OpenShift SDK to build the operator image and pushed the image to a container registry. And to deploy the Operator, all the Kubernetes artifacts in the deploy folder were created in the target cluster using razee. After the Operator is up and running, we communicate with the Kubernetes API to create the RStudio application instances.

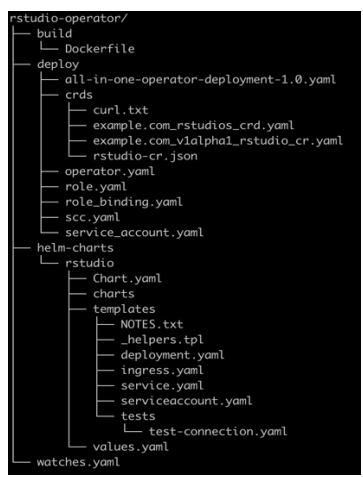

Figure 17: Directory structure showing the Open-source RStudio Operator development files

Many services in the Watson SaaS family are built on top of open-source products, such as IBM Watson Notebook, IBM Watson RStudio. They can be deployed in a similar way as their Open-source counterparts. This can be achieved by developing Watson Studio Kubernetes Operators for Notebook and RStudio and deployed using Razee. The IBM Blockchain Platform offered on IBM cloud currently uses Kubernetes operator to deploy the Blockchain Platform and our deployment model can be easily extended to it as well. We believe if there is a need to scale it to thousands of clusters, Razee can be deployed in concert with an operator to scale.

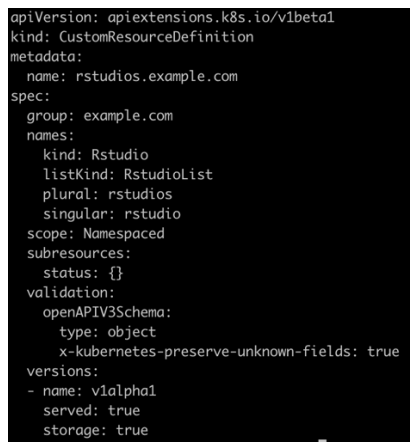

Figure 18: Custom Resource Definition of Open-source RStudio application

## 4 Discussion and Analysis

### 4.1 Comparison of Razee with Jenkins

After experimenting with Razee, we compare it with another widely used open-source CICD tool Jenkins.

| Feature | Razee | Jenkins Classic |
| --- | --- | --- |
| Continuous Integration | Not applicable, but can be integrated with any other CI tool like Tekton, Jenkins, etc., | Supported |
| Continuous Deployment | Supported | Supported |
| Environments | Works only with Kubernetes resources | Works with Kubernetes and non-Kubernetes environments |
| Deployment method | Pull based as the deployment job will run on the target cluster | Push based as the deployment job runs on the central Jenkins server. |
| Scalability | Comparatively highly scalable when dealing with 100s or 1000s of clusters | Comparatively less scalable |
| **Ease of configuration and setup environment** | Difficult to configure | Typically, moderately difficult |
| **Performance** | Pull based deployment model makes it high performing. Time to roll out Kubernetes resources in one cluster is the same as rolling out on 1000s of clusters | Comparatively less performing |
| **Integration with other tools** | Integrates very well with any CI tools | Very flexible |
| **Visualize Inventory and view deployed version history** | Supported | Not supported |
| **Continuous Monitoring of deployed resources** | Supported using Watchkeeper | Not supported |
| **Ease of analyzing the deployment errors alerts** | Razee maintains the inventory of deployment states. With the help of intelligent filters and alerting mechanisms, razee makes it easier to spot issues | Issues can be identified from deployment logs. As it doesn't maintain the inventory of deployment states, it is difficult to spot issues after the deployment is completed. |
| **Deployment templates** | Needs additional configuration such as Configmap to pass the deployment variables. | Templates can be created, and deployment variables can be passed dynamically for the deployment job. |

Table 1: Razee vs Jenkins comparison

### 4.2 Benefits of Razee and Kubernetes operators

We show the comparison with another popular CICD tool Jenkins. Our conclusion based on this table is that Razee along with Kubernetes Operators is better than Jenkins. Although the initial set up is complex, with several moving parts, it is more scalable than Jenkins. We think Jenkins is good for a small-scale project.

Razee will deliver Kubernetes resources to your clusters. In large-scale deployments, Razee becomes more beneficial because it allows you to deploy to many clusters at once.

The Kubernetes operators act more like Site Reliability Engineers if something goes wrong with the deployed operator or with the instance of the operator, it can bring them back to the functioning state. Kubernetes Operators extends numerous benefits to the CICD process. Kubernetes Operator knows how to create the entire Kubernetes stack from the CRDs. Instead of having Razee delivering all of the Kubernetes resources in yaml files to send to the cluster, the yaml file from Razee can be the CRD object, which is usually much smaller. The Kubernetes operators on the target clusters will then create an entire stack based on the CRD object. In addition, Kubernetes operators will handle Day 2 operations. The combination of Razee to deliver yaml files to thousands of clusters and Kubernetes Operators to create and manage resources automatically makes this solution scalable.

For example, let's say we want to deliver a stack of Kubernetes resources to thousands of clusters. The stack has Kubernetes services, deployments, ingress rules, Persistent Volumes, and Persistent Volume Claims. Without Kubernetes operators, we'll need to define all of these resources in yaml files and send them to the clusters. With Kubernetes operators, we only need to create the Custom Resource object and deliver this object to the cluster. The operator processes this Custom Resource object and creates all of these Kubernetes resources.

Razee and Kubernetes Operators have their strengths. We believe the combination of Razee to deliver yaml files to thousands of clusters and Kubernetes Operators to create and manage resources automatically will help achieving operational scalability.

## 5 Conclusion

In this paper, we demonstrated the use of IBM open-source project Razee as a scalable continuous deployment framework to deploy open-source RStudio and Nginx Operators and applications using the Operator. We also discussed on how the same deployment model can be extended to complex applications like IBM Blockchain Platform, IBM Watson Notebook, and RStudio products such as Notebook and RStudio and the use of Operators to perform application life cycle management. We also compared the advantages of Razee over Jenkins in terms of scalability and performance and we assert that using Razee in conjunction with Kubernetes Operators simplifies application life cycle management and increases scalability.

### ACKNOWLEDGMENTS
We would like to thank Michael McKay (IBM), Paul Portela (IBM) for their support and invaluable insights on the topics of Razee and Jenkins respectively.